\documentclass[a4paper,11pt]{article}
\pdfoutput=1 % if your are submitting a pdflatex (i.e. if you have
\usepackage{jinstpub} % for details on the use of the package, please
\usepackage{lineno}
\usepackage{siunitx}
\usepackage{eurosym}

\title{\boldmath Study of Spatial Resolution of Muon Hodoscopes for Muography Applications in Geophysics}

\author[a,b,c,d]{R. Calder\'{o}n-Ardila$^1$,
\note{Corresponding author.}
}
\author[a,d]{A. Almela,}
\author[b,e]{M. G\'{o}mez-Berisso,}
\author[a]{A. Sedoski,}
\author[a]{C. Varela,}
\author[b,f]{A. Vesga-Ram\'{i}rez,}
\author[a,b,e]{H. Asorey,}

\affiliation[a]{Instituto de Tecnolog\'{i}as en Detecci\'{o}n y Astropart\'{i}culas, Centro At\'{o}mico Constituyentes, Buenos Aires, Argentina.}
\affiliation[b]{Consejo Nacional de Investigaciones Cient\'{i}ficas y T\'{e}cnicas, Argentina.}
\affiliation[c]{Instituto SABATO, Universidad Nacional de San Mart\'{i}n, Centro At\'{o}mico Constituyentes, Buenos Aires, Argentina}
\affiliation[d]{Universidad Tecn\'{o}logica Nacional, Facultad Regional Buenos Aires, Argentina}
\affiliation[e]{Centro At\'{o}mico Bariloche and Instituto Balseiro, Comisi\'{o}n Nacional de Energ\'{i}a At\'{o}mica and Universidad Nacional de Cuyo, San Carlos de Bariloche, Argentina}
\affiliation[f]{International Center for Earth Sciences, Comisi\'{o}n Nacional de Energ\'{i}a At\'{o}mica, Buenos Aires, Argentina.}

\emailAdd{rolando.calderon@iteda.cnea.gov.ar}

\abstract{
Muon radiography, also known as muography, is a non-destructive geophysical technique for the study of the internal structure of large objects such as volcanoes.
This is possible by constructing an image based on the differential absorption of the directional flux of high-energy atmospheric muons produced during the interaction of cosmic rays with the atmosphere. So this no other source of radiation is required for this technique.
Many muon telescopes are being built with crossed scintillator bars and so, the resolution of each panel is essentially given by the total surface of the bar crossings.
Enhancing the resolution may require covering the same area with smaller scintillator bars, which adds costs and build complexity as more scintillators and fibers are required.
More channels also require more acquisition electronics which have to be synchronized, increasing the complexity of the system, with associated operating issues and the final cost.
In this work, we propose a novel analysis approach to obtain a reliable sub-pixel resolution,
by measuring and comparing the average signals measured at each end of the scintillation bar.
This analysis approach achieves sub-pixel resolutions, augmenting the spatial resolutions of existing designs.
To study the feasibility of this technique we designed a laboratory setup, to emulate muon light pulses with a pulsed laser light located at different points on optical wavelength shifter fiber. By doing this we measured an increase in the spatial resolution when compared with traditional systems.
These results enable the design of new prototypes for the muography of natural and artificial structures of strategic interest.
We are currently assembling a prototype detector that will use this methodology.  For its application, we have selected two locations for Copahue and Peteroa volcanoes in Argentina, where the detectors are planned to be installed. These active volcanoes present permanent volcanic gas emissions from a well-developed volcanic hydrothermal system, which make them suitable for this methodology. }


\arxivnumber{2006.03165} % only if you have one

\begin{document}
\maketitle
\flushbottom

%---- SECTION 1 -----

\section{Introduction}\label{sec:introduction}
%\linenumbers
Research in Geophysics is generally carried out with the help of different types of detectors or instruments.
These detectors correspond to the different methods used in geophysical prospecting like geoelectric, ground-penetrating radar, gravimetric, magnetotelluric, and seismic.
To study complex geological structures like volcanoes, glaciers or natural dams represents great difficulty because the different conventional geophysical methods have limitations.
Two of main limitations are resolution with depth\,\cite{jourde2015improvement,lelievre2019joint} and the danger derived from invasive techniques.  

During the last twenty years, the technique of muon tomography or muography has been developed to construct a digital image in terms of the contrast of densities product of the absorption of atmospheric muons. Muons ($\mu$) are secondary particles created in the atmosphere from primary cosmic rays and have been used in tomography techniques because of their great power of penetration into matter\,\cite{bugaev1998atmospheric}. In the case of atmospheric muons the flux is greater than that of other secondary particles such as protons or pions.

To study volcanic structures using the muography technique\,\cite{alvarez1970search, borozdin2003radiographic} an instrument is required that can detect muons. So that, for given exposure time, an image of the dome or volcanic structure can be constructed in terms of density contrast. Conceptually it is similar to an X-ray, but with atmospheric muons as the radiation source\,\cite{bonechi2020atmospheric}. The accuracy of the obtained image depends mainly on the spatial resolution of the instrument and the exposure time. In Latin America detectors have been developed for this purpose, such as the study of the internal structure of the Pyramid of the Sun in Mexico\,\cite{menchaca2014using}. In Colombia some projects for Volcano Muography like\,\citep{2017RMxAC..49...54A, vesga2020muon} and to determinate opacity in the Monserrate Hill\,\cite{parra2019estimation}.
Some designs are made up of arrays of plastic scintillators, a material capable of producing luminescence when a charged particle passes through them.
For most plastic scintillator designs, the spatial resolution depends on the total area where scintillators cross each other.

Recent research in muograph construction mentioned above motivate this work to study a possible improvement of the spatial resolution in plastic-scintillator based muographs\,\cite{platino2011amiga}. This research is the result of a technology transference from the experience gained from the AMIGA (Auger Muons and Infill for the Ground Array) project\,\citep{aab2016prototype} at the Pierre Auger Observatory in Malarg\"ue-Mendoza, Argentina; which developed and built detectors capable of counting atmospheric muons to study the muonic component of cosmic ray showers.
In our case, the first study targets for these prototypes will be the Copahue and Peteroa volcanoes  located in the Southern Volcanic Zone (SVZ) of the Andean range, on the border between Chile and Argentina (figure\,\ref{motivation}).

%---- SECTION 2 -----
\section{The Muon Radiography}
\subsection{Muon imaging}

Muons are elementary particles, have high energy, are similar to the electron but approximately 207 times your mass. The 2.2 $\mu s$ lifetime of muons is much longer than many other subatomic particles. This flux about the near-surface of Earth is constant, natural, abundant, and free to use. Additionally, muons are produced in the Earth’s atmosphere with known angular and energy distributions.
Muon flux is around 1 muon/cm$^{2}$ per minute, or 150 - 200 muon/m$^{2}$ per second, at sea level. Can travel through the large thickness (up to kilometers) of rock, with an energy loss mainly depending on the amount of matter crossed\,\citep{carbone2013experiment}. By measuring the absorption of muons along different paths through a solid body is possible to deduce the density distribution inside targets.
The differential flux of incident cosmic muons usually is given in ($\mathrm{cm}^{-2}~\mathrm{sr}^{-1}~\mathrm{s}^{-1}$ GeV$^{-1}$) and depends on the measurement site and the zenith angles. Also can be determined through either Monte Carlo simulation codes\,\citep{agostinelli2003geant4, athanassas2020simulation} or calculated from the analytical model modified through the empirical parameterization\,\citep{bugaev1998atmospheric,reyna2006simple,kudryavtsev2009muon}. Muon imaging is obtained through the comparison between the theoretical flux and observed integrated muon flux after the target. The theorical integrated muon flux is calculated assuming a homogeneous target  using the theoretical incident muon flux, open sky conditions, and relevant topographic information.

\subsection{Application to geological structures}

The muography technique has been successfully used in different fields like civil engineering\,\citep{guardincerri2016imaging,saracino2019applications}, archaeology\,\citep{menichelli2007scintillating,morishima2017discovery} and nuclear safety\,\citep{Ambrosino_2015,mahon2019first}. Muography in geoscience has many applications include prospecting \,\citep{schouten2019muon,bonneville2019borehole}, imaging of underground structures\,\citep{saracino2019applications} and, an interesting and important application today, the monitoring of carbon capture storage sites\,\citep{kudryavtsev2012monitoring,gluyas2019passive}. However, the most famous application in geosciences is the imaging of the inside of volcanoes. Is known as muon radiography in 2D or muon tomography in 3D.

Volcano muography is a technique that uses the directional flux of atmospheric muons to obtain a mass density image of a volcanic structure\,\citep{tanaka2009cosmic,LesparreMuon,tanaka2019japanese}.
The flux of muons has a dependence on the zenith angle and the mass traversed by the muons, to shows an illustrative image of a muograph setup for measuring a volcanic dome and the resulting estimation for the image, see\,\citep{moss2018muon,rodriguez2018minimute, Pe_a_Rodr_guez_2020,vesga2020muon}.

\begin{figure}[h!]
\centering
    \includegraphics[scale=0.49]{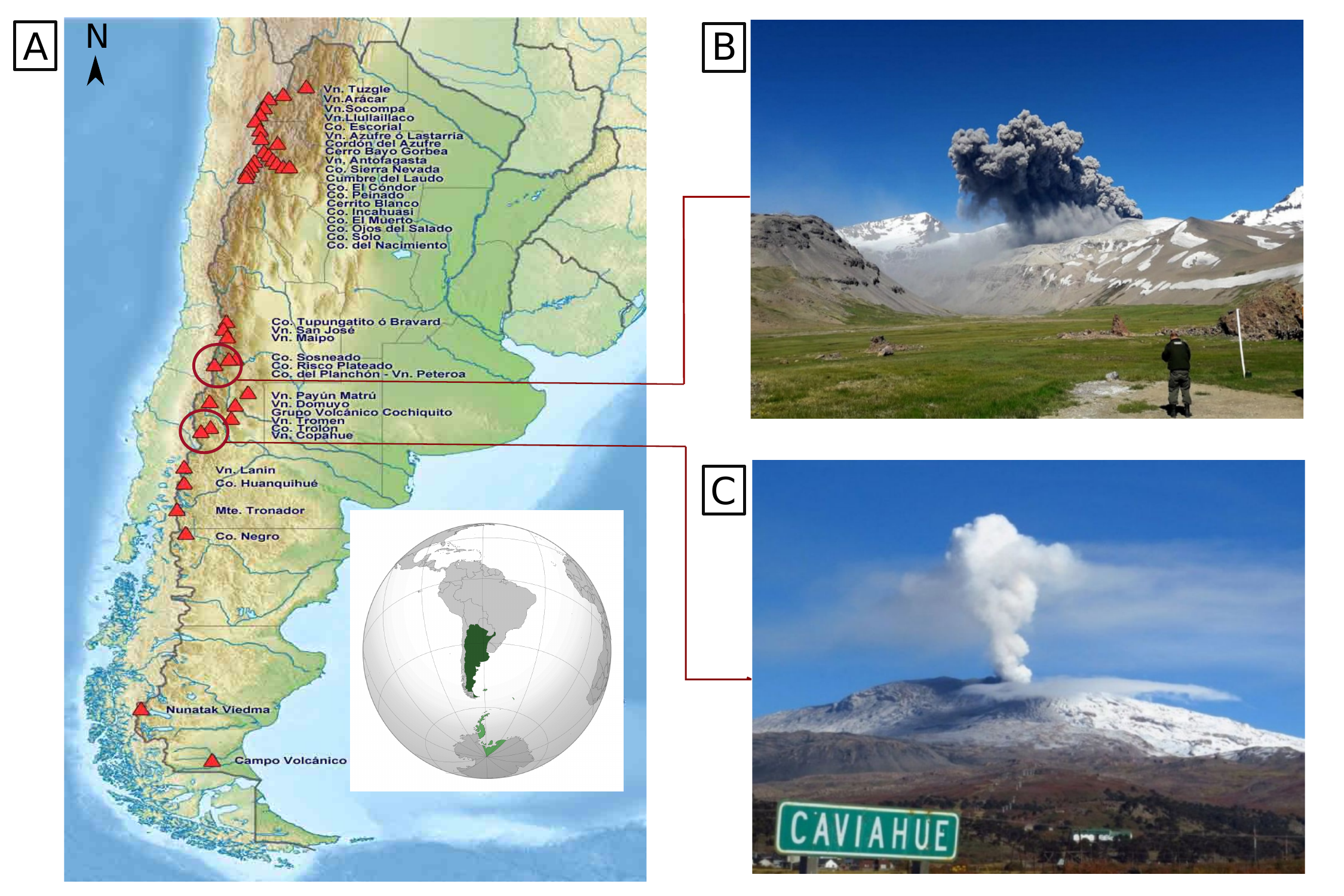}
  \caption{A) Schematic map showing volcanoes in Argentina, located in the Southern Andean Volcanic zone. This image was taken from (Source: GESVA-FCEN. UBA). One of the main motivations for the development of muography detectors in Argentina is the study of active volcanoes in the region, such as: B) The Planchón-Peteroa Volcano, located in the Transitional Southern Volcanic Zone of the Andean Ridge, and C) Copahue volcano located in the southwestern part of Caviahue Caldera.}
\label{motivation}
\end{figure}

\begin{figure}[h!]
\centering
    \includegraphics[scale=0.47]{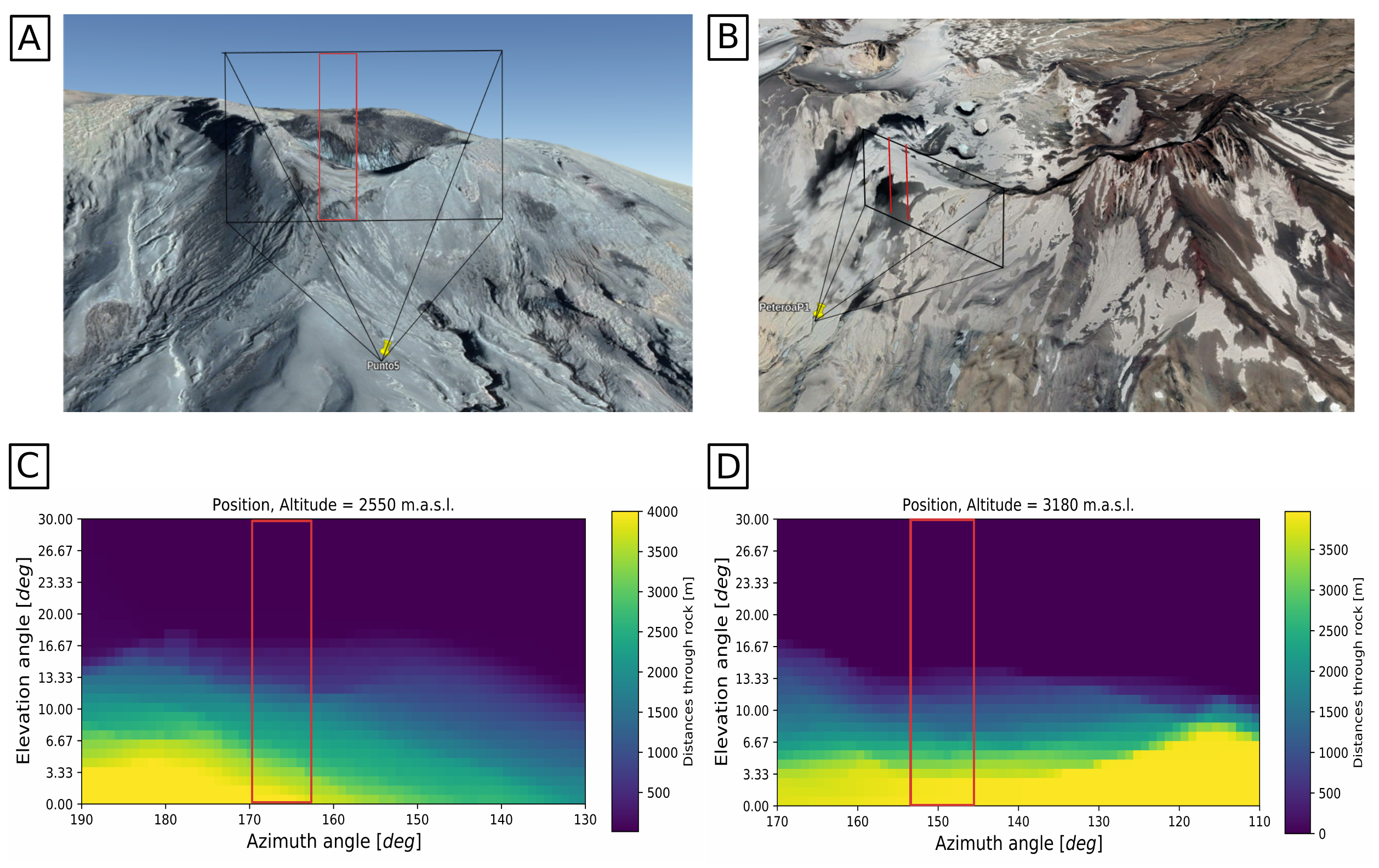}
  \caption{A) A representative image using Google Earth where We locate the reference point near the crater lake at Copahue volcano in Argentina with coordinates \SI{37}{\degree}\SI{85}{\arcmin}S, \SI{71}{\degree}\SI{15}{\arcmin}W. C) For the reference point, we present an image constructed by projecting rays through the dome of the volcano to estimate the length of the distance in the rock. This allows us to evaluate different detector location points depending on the thickness of rock traversed. B) A representative image using Google Earth for a reference point near the crater in the Planchón-Peteroa volcano in Argentina with coordinates \SI{35}{\degree}\SI{24}{\arcmin}S, \SI{70}{\degree}\SI{54}{\arcmin}W. D) For the reference point in Peteroa, We present an image constructed by projecting rays through the dome of the volcano to estimate the length of the distance in the rock. In all images, the red box represents the zone where density variations are expected to be seen due to volcanoes activity, see\,\citep{agusto2017crater,lamberti2021soil}.}
\label{modelo_rayos}
\end{figure}

By knowing the flux and morphology of the geological target it is possible to produce an image from the correlation between the muon count in each direction made with a muon telescope and the distances crossed inside the object\,\citep{rosas2018muon,varga2020tracking}.
In this way, it is possible to calculate the average opacity along the muon trajectories in the material traversed and to determine the density contrast in the final image.
Additionally, these calculations can be compared with simulations and synthetic data obtained by modeling the target\,\citep{vasquez2020simulated} and the muon flux\,\citep{AsoreyNunezSuarez2018,sarmiento2019modeling}.

The  muography technique has been used on some volcanoes around the world, such as Mt. Vesuvius by the Mu-RAY project\,\citep{ambrosi2011mu}.
This project built a muon telescope using three panels consisting of plastic scintillators with a triangular cross-section, coupled to optical fibers with silicon photomultipliers.
It has also been used in Japan where it has been successfully deployed on volcanoes such as Mt. Satsuma-Iwojima\,\citep{tanaka2009cosmic}, Mt. Sukuba\,\citep{nagamine1995method} and Mt. Asama\,\citep{tanaka2008radiographic}, studied using photographic emulsion.

In Italy, muography has been used as a complementary technique to study Mt. Etna\,\citep{carbone2013experiment}, but the morphology of the volcano hampered the data analysis. Authors report differences between synthetic models and  muon flux measurements through the target. The authors determine that the discrepancy is due to background flux from low-energy particles that are counted as muons but come from particle showers that occurred between the detector and the target. To solve this problem, they created a background noise model and used it to subtract from the measured signals. Finally, the results are compared with the corresponding synthetic flux.

In Colombia, the MuTe telescope\,\citep{vesga2020muon,vasquez2020simulated,vesga2021simulated} uses rectangular cross-section bars made of organic scintillator, wavelength shift optical fibers, photomultipliers, and acquisition electronics.
It is a hodoscope composed of arrays of scintillator bars called panels, each consisting of ${ N }_{ i }\times{ N }_{ j }$ bars.
These panels are attached to a mechanical structure, setting up the hodoscope.In this telescope for example, the resolution of the panels is determined by the number of crosses of the bars or pixels of the hodoscope.

To obtain a muography image, first is required to know a priori the properties of the object of study, for example in the case of a volcanic dome it is important to know its morphology and rock composition. Also, it's required to know the distances that the muons cross in rock within the field of vision of the detector.
For this, it is preferably to know the topography of the geological object  with a resolution of \SI{30}{\meter} at least. Knowing this information it is possible to generate muon flux synthetic data because the stopping power of the muons in rock is known.  
For example, if the target structure is bigger than \SI{2000}{\meter}, it will require a longer detector exposure time because most of the muons will be attenuated.
In all cases, the properties of rocks and the muons with your spectrum must be taken into account. For a more complete understanding of the possibilities and limits of the technique using muons, see\,\citep{LesparreMuon}.

After forward modeling, the directional muon flux is acquired with the instrument,
Then, flux attenuation is estimated in terms of opacity and finally a geophysical inversion can be made using the models and data obtained. Several examples of this methodology are presented in\,\citep{barnoud2019bayesian, benton2020optimizing,vesga2021simulated} where authors calculated the average density ($\tilde{\rho}$), measurement for each pixel as a function of the number of muons detected.

\begin{figure}
\centering
    \includegraphics[scale=0.35]{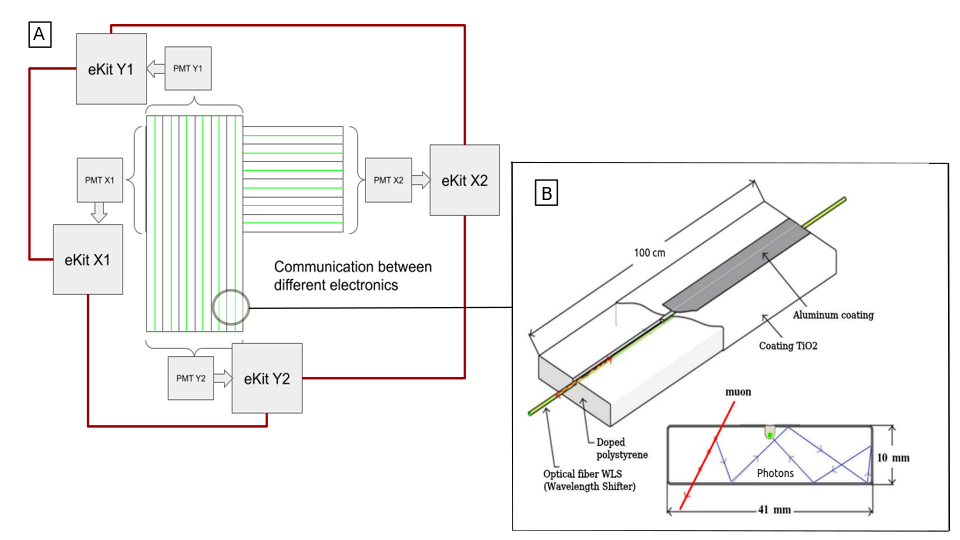}
  \caption{A) Prototype schematic (not to scale), has synchronized detection at the ends of each bar and a modular design to change spatial resolution. eKit refer to adquisition electronic system, every bar is similar~\citep{PHDAlmela}. B) Scintillator bar cross-section for each bar, for each area where there are crossings of the scintillator bars, we can identify $(2N_x~-~1)(2N_y~-~1)$ different particle trajectories, $r_{m,n}$, shared by the same relative position, $m~=~i-k$ and $n~=~j-l$, see Refs. \citep{LesparreMuon}.}
\label{modelo_panel}
\end{figure}

This work focuses on studying the improvement of the spatial resolution of this type of panels by performing a double detection for each scintillator bar, to determine the position of the bar where muons crossed, signal strength is measured at each end of the scintillator fiber.
The methodology proposed in this work will be applied to a muography prototype to be developed and built at Instituto de Tecnolog\'ias en Detecci\'on y Astropart\'iculas (ITeDA)\footnote{Instituto ITeDA \url{http://www.iteda.cnea.gov.ar/}}.
Our prototype has modules X-Y  arrays of ($12~\times~12$) or ($24~\times~24$) plastic scintillating strips with rectangular profile ($100$~cm$~\times~4$~cm$~\times~1$~cm). Each array has $144$ or $576$~physical pixels, giving a surface detection area up to $10000$~cm$^2$. The main idea for our prototype is to have a modular arrangement that can be configured (see figure~\ref{modelo_panel}). 

The candidates for field acquisition are the craters of Copahue and Peteroa volcanoes (figure~\ref{motivation}). These active volcanoes present permanent volcanic gas emissions from a constrained volcanic hydrothermal system at the summit\,\citep{agusto2017crater,lamberti2021soil}.  According to the preliminary models and simulations (figure~\ref{modelo_rayos}), these characteristics develop an efficient rheology contrast in the crater areas that makes them suitable for applying this methodology.

%---- SECTION 3 -----
\section{Instrumentation}\label{sec:scint}

The prototype muograph that motivated this work will be composed of plastic scintillator bars doped with 1 \% of PPO [2,5-diphenyloxazole] and 0.03 of the POPOP [1,4-bis(5-phenyloxazole-2-yl)benzene], with a $\textrm{TiO}_{\textrm{2}}$ coating\,\citep{pla2001low}, coupled at one side to an optical wavelength-shifter fiber. These fibers are the multi-layer reference BCF99-29AMC manufactured by Saint-Gobain, have a diameter of \SI{1.2}{\milli\metre} and produce a wavelength shift of the photons produced in the scintillator.

\begin{figure}[h!]
 \centering
    \includegraphics[scale=0.34]{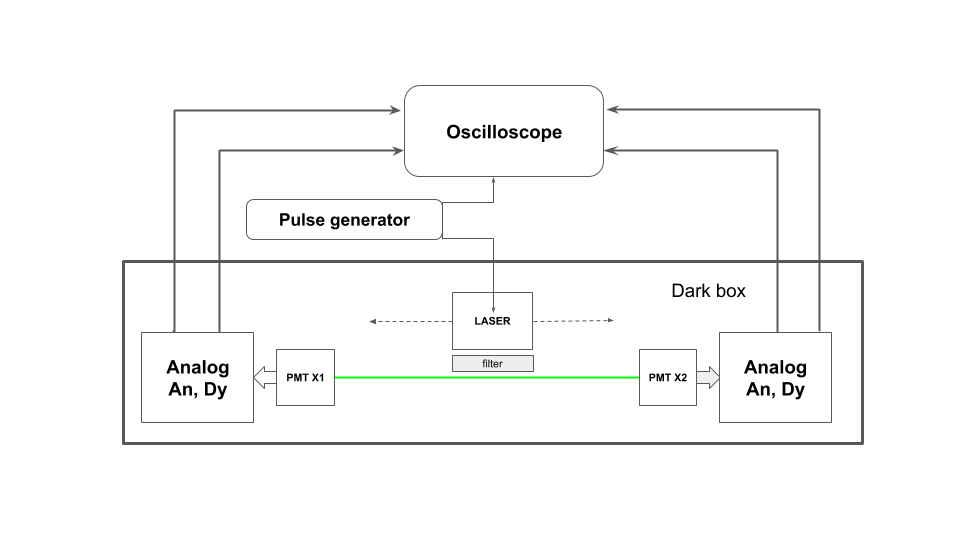} 
  \caption{Diagram representing the setup designed and built to determine the position of an event from the signal detected synchronously with two acquisition electronics and a pulse generator.
  Inside the dark box, the position of the laser represents the location of the artificially generated event.}
\label{esquema1}
\end{figure}

When a muon passes through the scintillator, a portion of the photons produced are adsorbed and re-emitted by the wavelength shifter (WLS) fibers to finally be guided to one channel of a H8804-b 64-anode photomultiplier manufactured by Hamamatsu.
The pulses generated at the PMT are saved, analyzed and identified as events.
This detection sequence has been tested and is used by the AMIGA project\,\citep{aab2016prototype} and the MuTe project\,\citep{Pe_a_Rodr_guez_2020,vesga2020muon,vasquez2020simulated}.

\begin{figure}[h!]
  \centering
    \includegraphics[scale=0.4]{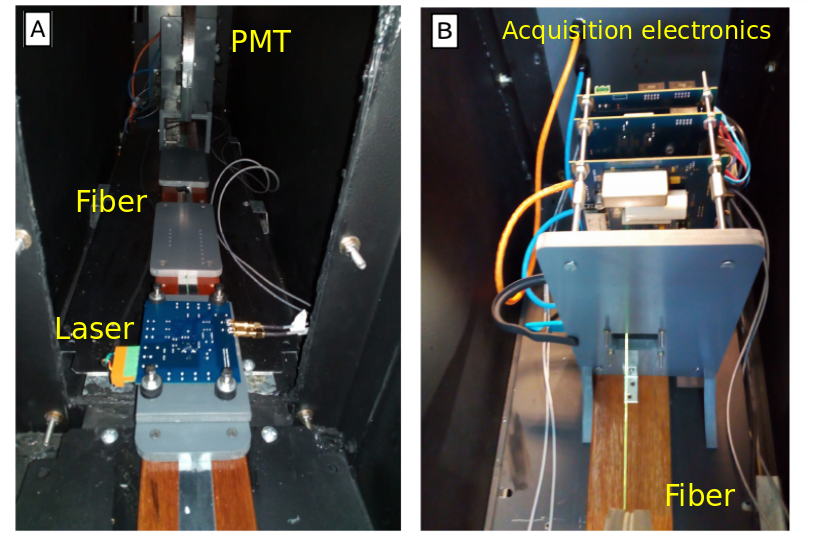}
      \caption{Dark box and setup built to perform the synchronous detection. The dark box was designed and built as a light insulator. A) Inside the dark box a WLS fiber was installed, with a device designed to install a pulsed laser light source at different positions. B) Each end of the WLS fiber is coupled to a PMT channel and acquisition electronics.}
  \label{Figura4}
\end{figure}

The charge of a PMT pulse is given mainly by the amount of photons sensed.
The light arriving to one end of the fiber depends on the light adsorbed by the fiber and the light attenuation, which is a function of the distance from the measurement point to the point of adsorption.
We characterized these effects with an experimental setup consisting of a WLS fiber coupled at each end to a PMT channel and its associated acquisition electronics.
This setup allows the placement of a light source at different positions, see figure~\ref{Figura4}.
We chose a \SI{405}{\nano\metre} laser to act as a light source, which has a similar wavelength to the light emitted by the plastic scintillator.
This laser is triggered by a pulsed laser driver, driven by a pulse generator (Agilent N8042A) that can generate pulses with a similar duration to a typical muon pulse and provides synchronization to the acquisition electronics.
A diagram of this setup is shown in figure~\ref{esquema1}.
Fifty thousand samples were acquired for each position using an oscilloscope (LeCroy WaveSurfer 104MXs-B).

As described in section~\ref{sec:Acqui} The PMTs used for this characterization is calibrated, to obtain the expected charge for each photo-electron\,\citep{bellamy1993absolute}.
After both PMTs were calibrated, measurements were taken at 24 separate positions on a \SI{180}{\centi\metre} long fiber every \SI{5}{\centi\metre}.
In Our prototype the attenuation is very low because optical fiber has \SI{90}{\centi\metre} from middle position to  at each end. The laser source was suitable (optical filters) based on measurements witht muons, to \SI{1}{\metre} from PMT, the average of SPEs (Single Photo-electron) equivalent is $\sim$16 SPE. This has been previously measured in the laboratory\,\citep{platino2011amiga}.

\begin{figure}
 \centering
  \includegraphics[scale=0.30]{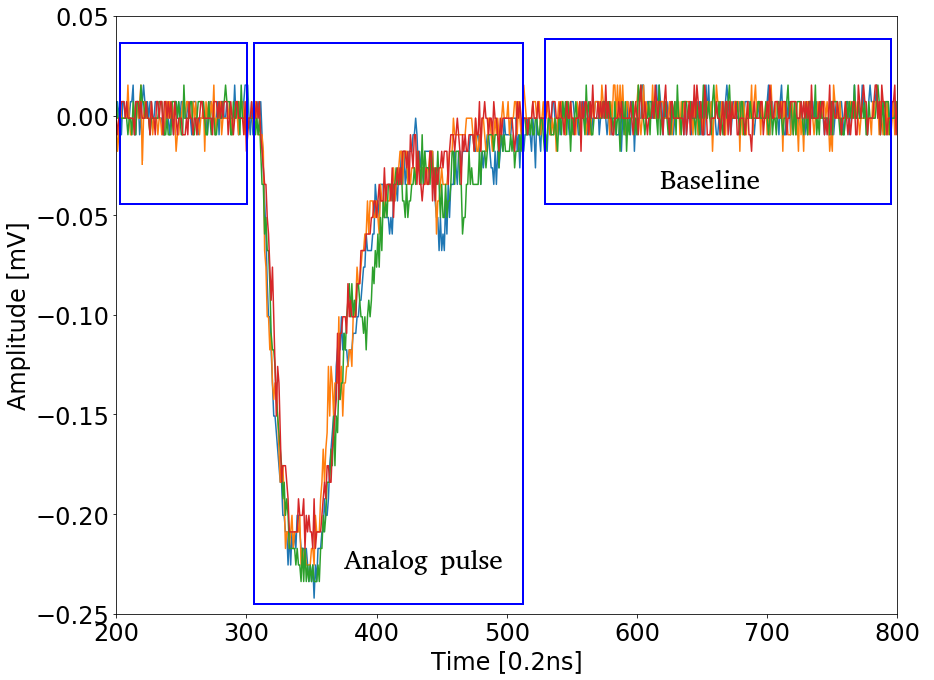} 
  \caption{Typical PMT pulses from a single anode, measured with a \SI{50}{\ohm} impedance.
  Each PMT has 64 outputs (``channels'') corresponding to the anodes.
  Each channel produces analog pulses as shown in this figure.
  For each anode, the pulse amplitude is always proportional to the number of impigning photons on the photocathode. 
  For the dynode, the pulse amplitude is proportional to the response from all anodes to the impigning photons.}
\label{PulseAnalogic}
\end{figure}

The schematic drawing of the setup is presented in figure \ref{esquema1}.
For each position, the histograms of the deposited charge at each end were obtained.
The charge was calculated from the oscilloscope traces by subtracting the baseline from the pulse and applying a point-to-point integration in a fixed time window corresponding to the area defined as the analog pulse (see figure \ref{PulseAnalogic}), so that each point will have a charge value ${ Q }_{ t }=\sum _{ a }^{ b }{ A(t)dt }$.

Then, based on these histograms, the number of photo-electrons detected was calculated as a function of the distance.
This function is the crucial part of the methodology on a muograph, it provides an accurate estimation of the impact point of the particle within the scintillator strip, enhancing the resolution of the device.

%---- SECTION 4 -----
\section{Data acquisition}\label{sec:Acqui}

The PMT used allows the acquisition of its 64 independent anodes and a common dynode, which integrates the charge collected by the anodes.
For calibration and measurement it was necessary to condition the light source by using Kodak optical filters.
For PMT calibration, filters were used to match the photo-electron count to the expected average at a given reference distance.
For data acquisition, this was done to simulate different muonic responses based on the Poisson probability distribution for the detection and quantum efficiency of the photo-cathode.

To calculate the mean of photon-electrons we fitted the acquired data with a PMT response model (adapted from\,\cite{bellamy1993absolute}).
With this model we obtained the best fit for low light emission conditions where most detected events correspond to a single photon.

The model used for the PMT response was augmented to obtain the charge ratio between an anode and the dynode.
This ratio is critical to the proposed mugraph prototype, as the AMIGA electronics does not allow the charge measurement of the anode.
As we are planning to operate the dectector by measuring the charge at the dynode, we discarded events where more than one anode detects signal. This means that the coincidence rate is low and only the dynodo signal will be used to locate the event. Given the expected muon flux for muography applications, this does not result in a significant event discard rate, and therefore does not increase the required exposure time.

\subsection{PMT Calibration}

The calibration method is based on the deconvolution of the pulse spectrum into a noise component and a pulse component, discriminated by number of photoelectrons sensed.
The characteristic parameters each number of photoelectrons detected are then also expressed as a function of the parameters for a single detected photoelectron.

An ideal PMT can be treated as an instrument consisting of two independent parts: a photodetector, where a photon flux is converted into photoelectrons, and an amplifier based on a dynode chain, which amplifies the initial charge emitted by the photocathode.

The flux of photons impigning on the PMT photocathode produces photoelectrons through the photoelectric effect.
For the photo conversion the $Q_E$ quantum efficiency is evaluated and described by:

\begin{equation}
\mu = \mu_{\rho} Q_E
\end{equation}

where $\mu$ is the average number of photoelectrons emitted and $\mu_{\rho}$ the average number of incident photons.
The number of photoelectrons emitted by the photocathode is a Poisson process, where $\mu$ is the average number of photoelectrons collected by the first dynode.
This model yields the equation:

\begin{equation}
P(n;\mu) = \frac{\mu^{n} e^{-\mu}}{n!}
\label{eqn:poisson}
\end{equation}

where $P(n; \mu)$ is the probability of observing $n$ photo-electrons when the average amount of photoelectrons detected per event is $\mu$.
Furthermore, when an average number of $\mu_{\rho}$ photons imping on the photocathode per event, the quantum efficiency of the photocathode is $Q_E$.
Thus, for a given PMT, the parameter $\mu$ characterizes the intensity of the light source, taking into account the quantum efficiency of the photocathode and the efficiency of the electron collection of the dynode system.

For the amplification response we model the dynode chain, where in each stage of the amplification cascade we have a Gaussian distribution\,\cite{chirikov2001precise}.
The total response of the amplification chain is the sum of the response of each dynode, giving the equation:

\begin{equation}
\label{eqn:gauss}
G(x) =   \frac{1}{\sigma\sqrt{2\pi}} e^{ - \frac{\left(x-q\right)^{2}}{2\sigma^{2}} }
\end{equation}

where $x$ is the charge received at the first dynode, $q$ is the average charge at the PMT output when an electron is picked up by the first dynode (single photo-electron, SPE) and $\sigma$ is the standard deviation of the entire amplification chain.

The charge distribution of a process initiated by $n$ photo-electrons is the convolution of $n$ processes with an SPE, therefore we can approximate the ideal response of the PMT without the noise as the convolution of the distributions \ref{eqn:poisson} and \ref{eqn:gauss}:

\begin{equation}
 S_{ideal}(x) = P(n_{SP};u) \otimes G_n(x)  
\end{equation}

obtaining an ideal charge model for the PMT:

 \begin{equation}
S_{ideal}(x) =  \sum_{n=0}^{\infty }  \frac{ u^{n} \cdot e^{-u} }{n!}. \frac{1}{\sigma \sqrt{2 \pi} }\cdot  e^{ - \frac{(x - n \cdot q)^{2}}{2 n\sigma^{2}} } 
\label{eqn:PMideal}
\end{equation}

This distribution alone does not represent the response of a real PMT due to device noise.
Background noise processes (such as  \textit{dark current}, \textit{afterpulse} and \textit{cross-talk}) can generate additional current signals that modify the output response.

These background processes can be divided into two groups, with different distribution functions: low charge processes and discrete processes.
Low-charge processes are responsible for non-zero current when there are not photo-electrons emitted from the photocathode.
This distribution corresponds to the pedestal and can be modelled with a Gaussian function.

\begin{equation}
\label{eqn:gaussR}
N_G(x) = \frac{1}{\sigma_{o}\sqrt{2\pi}} e^{ - \frac{\left(x-q_{o}\right)^{2}}{2\sigma_{o}^{2}} }
\end{equation}

The discrete processes can be thermo-emission (electrons detached from the photocathode without an impigning photon) and noise initiated by the measured light with probability other than zero. These processes accompany the measured signal and can be modeled with an exponential distribution.

\begin{equation}
\label{eqn:expR}
    N_E(x) =   \lambda e^{x\lambda}
\end{equation}

If call $w$ to the probability that a measured signal is accompanied by a background process, the background noise processes can be parameterized as:
\begin{equation}
    \label{eqn:fondo}
    N(x) =  \frac{(1-w)}{\sigma_{o}\sqrt{2\pi}} e^{ - \frac{(x-q_{o})^{2}}{2\sigma_{o}^{2}} }+ w \lambda e^{x\lambda}
\end{equation}

The first term corresponds to the discrete processes.
The second term corresponds to low charge background processes (pedestal).
For small $\sigma_{o}$( $\ll 1 / \lambda)$ (good baseline), the convolution of a Gaussian with an exponential function is reduced to a pure exponential function.

Using the spectrum for an ideal PMT from equation \ref{eqn:PMideal} and the background charge distribution in equation \ref{eqn:fondo}, the response spectrum for a realistic PMT can be obtained as the convolution of both distributions:

\begin{equation}
S_{real}(x) = S_{ideal}(x) \otimes N(x) 
\label{eqn:real}
\end{equation}

The response function for a real PMT contains seven free parameters and was developed at~\citep{PHDAlmela}.
Two of them, $q_{o}$ and $\sigma_{o}$, define the pedestal and two others, $w$ and $\lambda$, describe the discrete background processes. Finally the parameters $q$, $\sigma$ and $\mu$ describe the spectrum of the real signal.
Of these three parameters, $\mu$ is proportional to the intensity of the light source, and $q$ and $\sigma$ characterize the amplification process of the PMT dynode system.
Finally, we use the $S_{real}(x)$ response model to compare and calibrate the response of the PMTs.

\begin{figure}
\centering
    \includegraphics[scale=0.35]{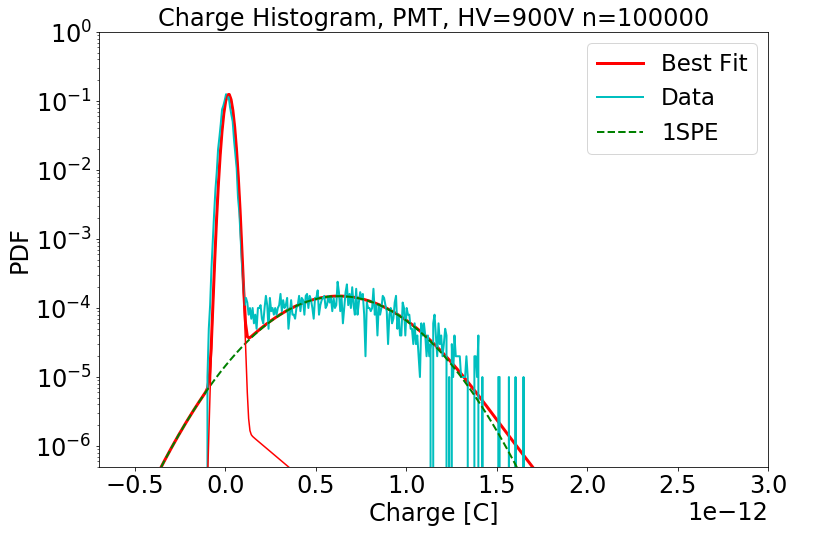} 
  \caption{Calibration histogram for a PMT.
  100.000 samples were taken in low light conditions for this purpose.
  The image contains two peaks: the first (leftmost) peak represents the most likely charge value for background noise, while the second (rightmost) peak represents the average charge for an SPE.
  }
\label{calibracion1}
\end{figure}

\begin{figure}
\centering
    \includegraphics[scale=0.23]{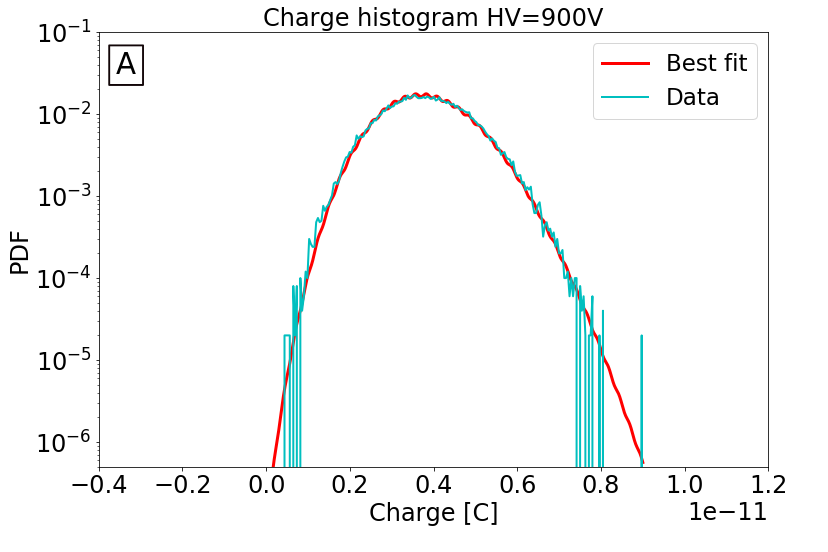}
    \includegraphics[scale=0.23]{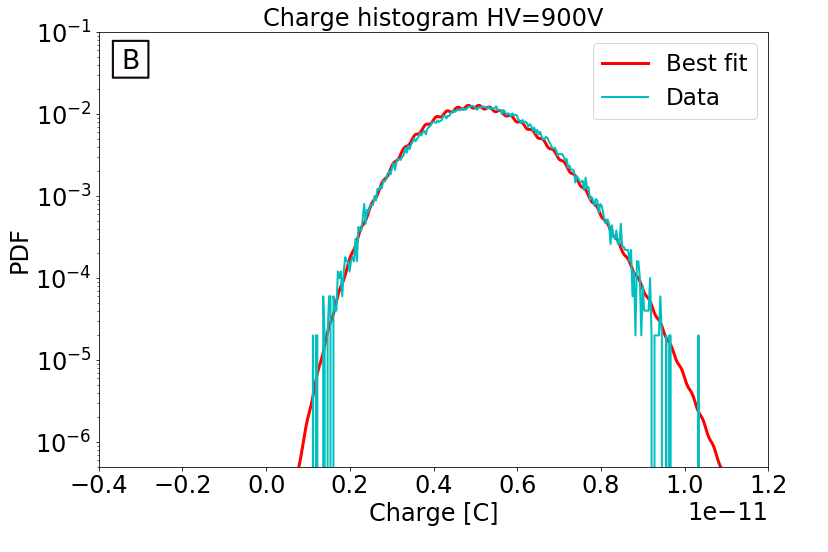}
  \caption{A) Histogram for an acquisition with an average of 14 SPEs, and  B) 20 SPEs.For each histogram, 50.000 acquisitions were considered. In both histograms the pedestal is not visible because at these light intensities and for this PMT most events detect at least one SPE.}
\label{calibracionPMT2}
\end{figure}

\subsection{Measurements using expected operation conditions}

After the PMTs were calibrated, we removed some of the light filters from the setup to simulate operating conditions and proceeded to measure pulses, moving the light source as described in \ref{sec:scint}.

\begin{figure}
 \centering
    \includegraphics[scale=0.40]{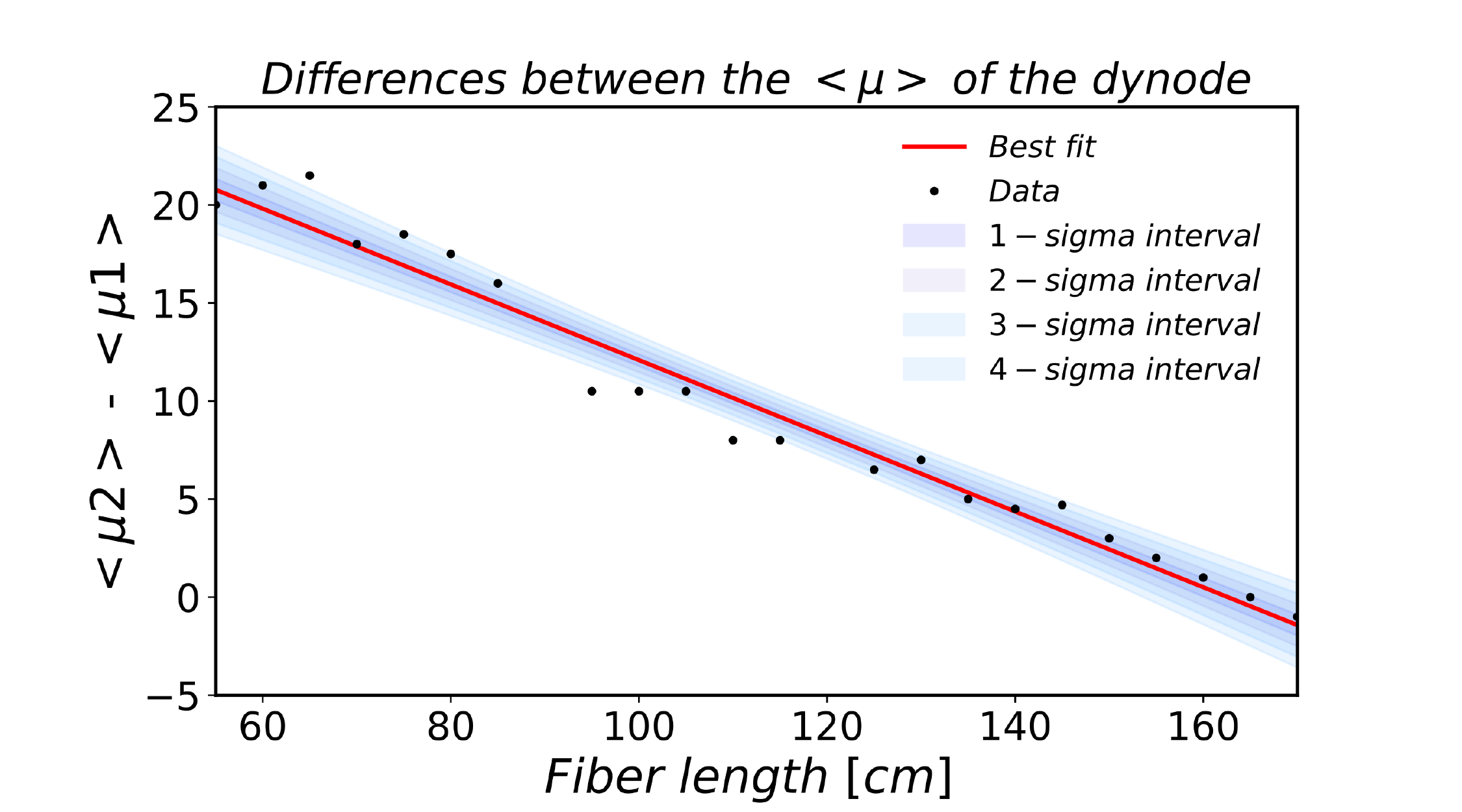} 
  \caption{After the calibration, the light source was moved lengthwise along the optical fiber.
  50.000 samples were taken at each light source position, and the $\mu$ was estimated for both electronics.
  Then we proceeded to subtract the $\mu$ calculated for each PMT at every position, then used these values to estimate the best fit and represent the confidence intervals in function of the $\sigma$ error. 
  In this picture we compare the measurements with the fit confidence intervals from 1-$\sigma$ to 4-$\sigma$.}
\label{interval-sigma}
\end{figure}

Thus we could establish the location within the bar according to the absolute difference in terms of $\mu$.
In each case $\mu$ was calculated from the fit values of the calibration for each PMT. 
At each position, 50.000 pulses were acquired, obtaining charge histograms as shown in figure \ref{calibracion1}.

\begin{figure}[h!]
 \centering
    \includegraphics[scale=0.32]{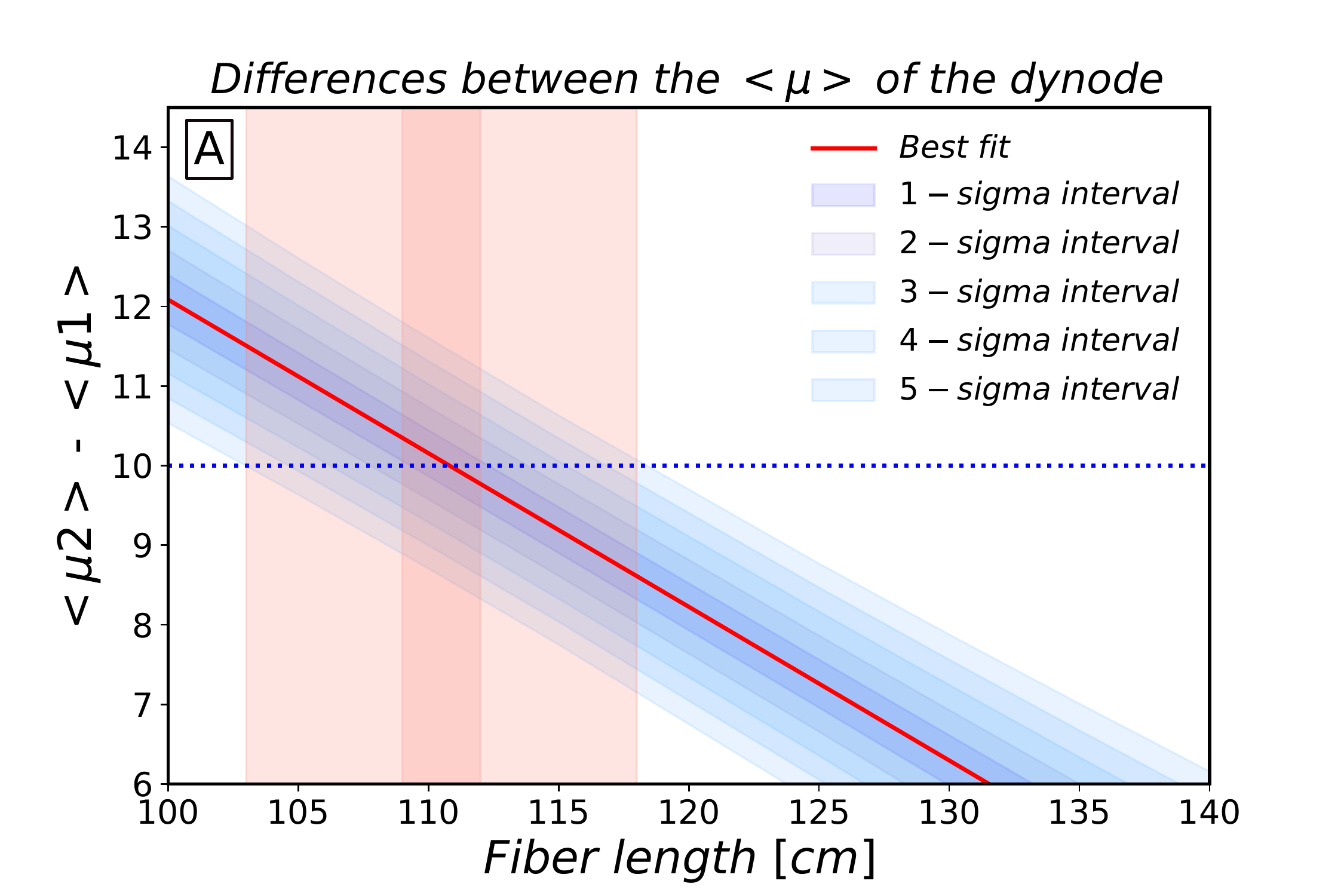}
    \includegraphics[scale=0.32]{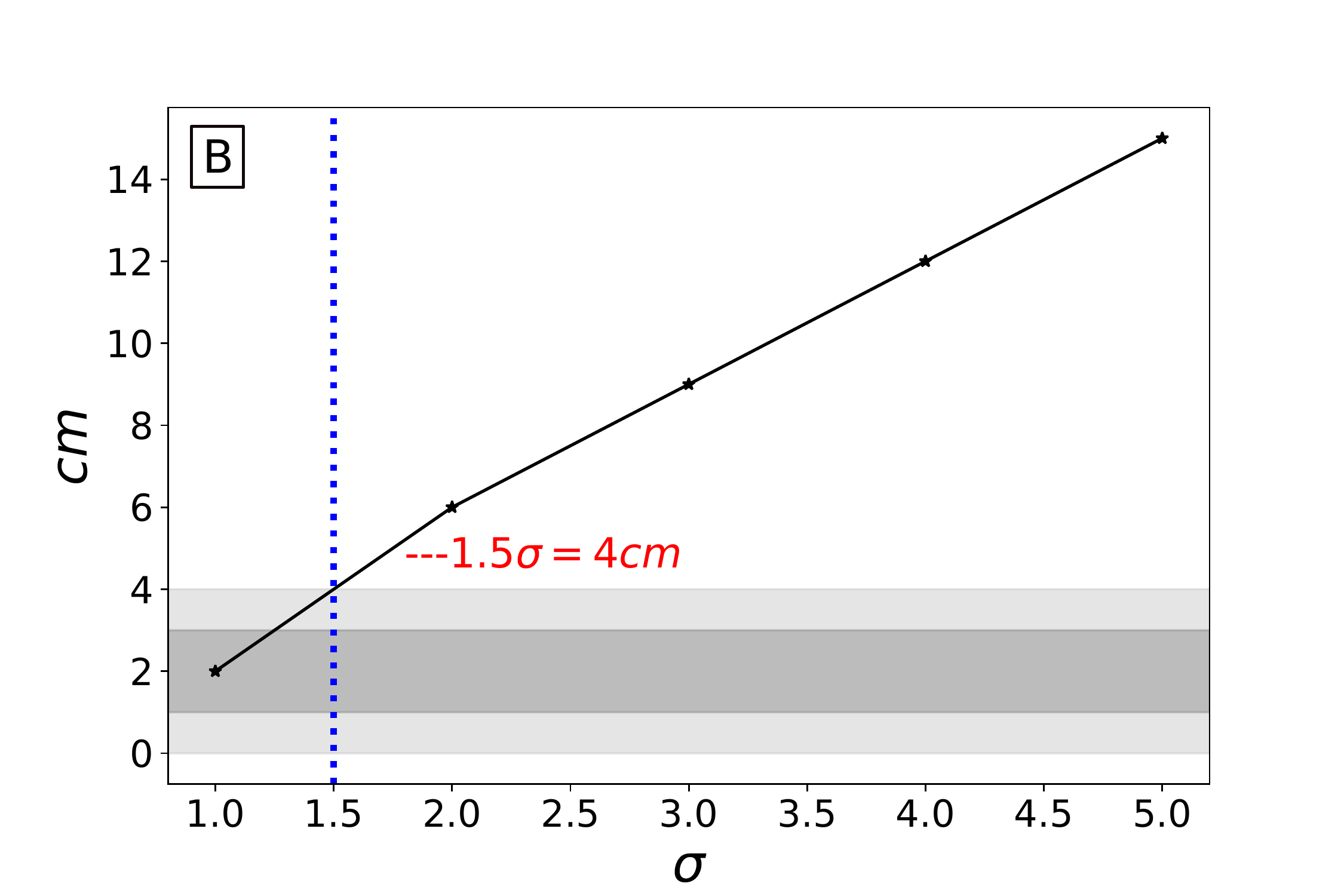}
  \caption{A) Projection of the confidence intervals as a function of the distance between the photocathode and the light source. B) Comparison of the uncertainty in terms of sigma $\sigma$ for different fiber lengths, using data from the previous plot. 1.5$\sigma$ represents the maximum acceptable uncertainty for a 4$\times$4 cm pixel.}
\label{final}
\end{figure}

Next, we subtract the average number of photoelectrons measured at each end for each position and present a graph where on the y-axis we have the subtraction of mu photoelectrons versus distance. Then using the error of the average measured value we calculate the covariance matrix with the best fit parameters and with them we estimate the sigma $\sigma$ confidence intervals as shown in figure\,\ref{interval-sigma} and \ref{final}. The goal is that the  determination of position does not depend only on the crossed pixels, but also on the subtraction of the signals based on the deposited charge\,\citep{dossi2000methods}. Where mu~($\mu$) representing the average number of equivalent photoelectrons inside the model\,\citep{bellamy1993absolute}. Thus, we have additional information to make an offline adjustment and improve the physical resolution without adding panels.

%---- SECTION 5 -----
\section{Discussion: Muography and conventional geophysical methods}

Does muography replace traditional geophysical methods? Not really, but it can certainly help and be advantageous near the surface. First, it must be remembered that determining the density distribution within geological structures is of great importance in the earth sciences. In particular volcanic activity involves several phenomena, such as the complex geometry of gas or liquid conduits, in some cases gas-liquid flows.
The geometry of the upper dome system of an active volcano is a key feature for modeling the processes that can lead to proximal activity.
Indeed, parameters such as maximum sustainable duct overpressure, magma ascent rate, gas volume fraction, and volumetric flow rate are related to the shape and size of the duct\,\citep{carbone2013experiment}.
This implies that detailed knowledge of the shallow underground structure is mandatory to forecast the occurrence of dangerous stages of activity and therefore could help mitigate volcanic hazards\,\citep{gilbert2008consequences}.

To investigate the underground structure of a volcano, different conventional techniques are used, which are sensitive to the physical properties of the underground rocks (like elastic, electrical properties and density). But are they sufficient?
Classical seismic approaches are based on the inversion of earthquake traveltime data to obtain a three-dimensional image of the structure using wave velocities\,\citep{haberland2009structure}. However, it has the problem of multiple reflections and hence a complex analysis in the case of the volcanic dome. Ambient noise interferometry is also increasingly used to image volcanic structures, see\,\citep{brenguier20073}.
This method is based on ambient seismic noise excited by shallow sources like oceanic microseisms or atmospheric disturbances and relies on the calculation of noise cross-correlations between different detectors. Subsurface distribution and electrical properties can be visualized through direct current electrical resistivity\,\citep{pessel2003multiscale} or low-frequency electromagnetic techniques, but these techniques exhibit rapid attenuation with depth.

Gravity measurements allow the calculation of Bouguer anomaly, which reflect anomalies in the subsurface density structure\,\citep{deroussi2009localization}. However, the correction can have a very large error at depth. For studies performed at local scales, from kilometer to decimeter, classical gravimetric methods are still the main approach used to retrieve the density distribution below ground.
Despite the vast improvements of gravimeters, either relative or absolute, the study of gravity remains a long, expensive and difficult task, especially at volcanoes and in rugged topography\,\citep{carbone2003combined}. Strong heterogeneities of the medium produce marked scattering/attenuation of the seismic wave and can lead to high resistivity/density contrasts, which makes the inversion of the data highly nonlinear in addition to the non-uniqueness that characterizes the gravimetric inverse problem.

Even with the numerous scanning techniques mentioned above, imaging volcanoes remains a challenge, latter because measurements must be made on active zones\,\citep{tiede2005modeling}, which implies a high level of risk.
Moreover, in most cases, the most established geophysical imaging techniques do not offer the spatial resolution needed to adequately characterize the shallow part at the summit of a volcano. Crater areas in active volcanoes are characterized by accumulation of unconsolidated materials of low density, particularly exposed to destabilization due to their low strength and high fluid content.
 
Destabilization of these vulnerable materials could trigger further building damage. Is for this reason why modeling the density structure of lava domes at the conduit affected by intense hydrothermal alteration is of paramount importance contributing with valuable information for hazard assessment. As aforementioned, evaluations from direct modeling and simulations point out that the crater areas of Copahue and Peteroa volcano show a structure according to these characteristics\,\citep{agusto2017crater,lamberti2021soil} and appropriated to apply this methodology. Here, muography allows exploration with a better spatial resolution and reduced ambiguity, and therefore improving the knowledge about their inner structure and their behavior.  

As mentioned in chapter 2, there are already several technological developments and local prototypes, it is a low cost, about $<$ 10K$\euro$/m$^2$, see Refs.\,\citep{bonechi2020atmospheric}, the implementation of the reconstruction algorithms is not complicated, muons can penetrate up to \SI{2000}{\metre}.
It can have image resolutions of between $>$10 mrad to $\sim$ 0.1 mrad, of course depending on the technology used and other parameters like distance from the detector to the dome\,\citep{bonomi2020applications}. This represents a spatial resolution of the order of meters on the object (volcanic dome).
Another interesting issue is the fact that the detector can be left measuring on-site (no continuous monitoring required). Exposure times is around several months for distances between \SI{100}{\metre} $\sim$ \SI{1500}{\metre} and a standard rock density  $\rho$ = 2.65 g.cm$^{-3}$ . In addition to the possibility of cross-checking data with other techniques, muography can be used as a complementary technique at the surface. Finally, as mentioned before, muography is a non-invasive technique and does not require additional sources of muon flux. Also, muons are safe, it is a natural source, abundant and free to use.

%---- SECTION 6 -----
\section{Conclusions}

A methodology and design modification to enhance the spatial resolution of a muograph was proposed.
An experimental setup using a movable controlled light source synchronized with acquisition electronics was built to test the feasibility of the proposed methodology.
This setup was used to calibrate PMTs using a PMT response model adapted to the proposed analysis methodology.

Using this setup with light intensities similar to expected as operating conditions, we could determine the position of the light source from the charge measured by the dynode of the PMTs at both ends of the fiber.
The confidence intervals of the measurements in operating conditions yielded favorable information about the uncertainty when determining the impact position using this strategy.
From these results we can estimate \SI{3}{\centi\meter} with a 1$\sigma$ of error with respect to the fit.
Since the muon flux is a known measurement, we can also estimate the exposure time to acquire enough events.
Therefore, this methodology is feasible.

Due to the measurement setup, no scintillator effects were considered. The experimental setup can be used to evaluate different optical fibres and light sensing devices. The methodology developed in this work will be implemented in our prototype in the craters of Copahue and Peteroa volcanoes, where the preliminary direct models and simulations suggest the hydrothermal system at the summit offers an efficient rheology contrast for applying this methodology. Thus, muography exploration will contribute to improving the knowledge of the inner structure and behaviour of these volcanoes, and therefore with valuable information for hazard assessment.

%---------------------------------------------------
\section*{Declaration of competing interest}

The authors declare that they have no known competing financial interests or  personal  relationships  that  could  have  appeared  to  influence  the  work reported in this paper.
%--------------------------------------------------------

\section*{Acknowledgments}

The autors are very thankful to the participating institutions, In addition this work is part of the PID-UTN5202 from Universidad Tecnológica Nacional and the SIIP-UNCuyoC035 from Universidad Nacional de Cuyo, additionally to Pierre Auger Collaboration from Pierre Auger Observatory for their support. Some results presented in this paper were carried out using the ITeDA computer cluster maintained by CNEA in Argentina.

%\nolinenumbers

%----------------------------------
%\section*{References}

%\bibliography{Biblio_ROCAL}

\begin{thebibliography}{10}

\bibitem{jourde2015improvement}
K.~Jourde, et al.
\newblock Improvement of density models of geological structures by fusion of gravity data and cosmic muon radiographies.
\newblock {\em Geoscientific Instrumentation, Methods and Data Systems}, 4(2):177--188, 2015.

\bibitem{lelievre2019joint}
Peter G.~Leli{\`e}vre, et al.
\newblock {Joint inversion methods with relative density offset correction for muon tomography and gravity data, with application to volcano imaging}.
\newblock {\em Geophysical Journal International}, 218(3):1685--1701, 2019.

\bibitem{bugaev1998atmospheric}
E.V.~Bugaev, et al.
\newblock Atmospheric muon flux at sea level, underground, and underwater.
\newblock {\em Physical Review D}, 58(5):054001, 1998.

\bibitem{alvarez1970search}
L.W.~Alvarez, et al.
\newblock Search for hidden chambers in the pyramids.
\newblock {\em Science}, 167(3919):832--839, 1970.

\bibitem{borozdin2003radiographic}
K.N.~Borozdin, et al.
\newblock Radiographic imaging with cosmic-ray muons.
\newblock In {\em Nature}, 422(6929):277. Nature Publishing Group, 2003.

\bibitem{bonechi2020atmospheric}
L.~Bonechi, et al.
\newblock Atmospheric muons as an imaging tool.
\newblock {\em Reviews in Physics}, 100038, Elsevier 2020.

\bibitem{menchaca2014using}
A.~Menchaca-Rocha, et al.
\newblock Using cosmic muons to search for cavities in the Pyramid of the Sun, Teotihuacan: preliminary results.
\newblock {\em PoS}, 012, SISSA 2014.

\bibitem{2017RMxAC..49...54A}
H.~Asorey, et al.
\newblock Astroparticle Techniques: Colombia Active Volcano Candidates for Muon Telescope Observation Sites.
\newblock {\em Revista Mexicana de Astronomia y Astrofisica Conference Series}, 49:54-, 2017.

\bibitem{vesga2020muon}
A.~Vesga-Ram{\'i}rez, et al.
\newblock Muon Tomography sites for Colombian volcanoes.
\newblock {\em Annals of Geophysics}, 63(6):661, 2020.

\bibitem{parra2019estimation}
J.S.~Useche-Parra and C.A.~\'{A}vila-Bernal.
\newblock Estimation of cosmic-muon flux attenuation by Monserrate Hill in Bogota.
\newblock {\em Journal of Instrumentation}, 14(02):P02015, IOP Publishing 2019.

\bibitem{platino2011amiga}
M.~Platino, et al.
\newblock AMIGA at the Auger Observatory: the scintillator module testing system.
\newblock {\em Journal of Instrumentation}, 6(06):P06006, IOP Publishing 2011.

\bibitem{aab2016prototype}
A.~Aab, et al.
\newblock Prototype muon detectors for the AMIGA component of the Pierre Auger Observatory.
\newblock {\em Journal of Instrumentation}, 11(02):P02012, IOP Publishing 2016.

\bibitem{carbone2013experiment}
J.~Carbone, et al.
\newblock An experiment of muon radiography at Mt {E}tna ({I}taly).
\newblock {\em Geophysical Journal International}, 196(2):633--643, Oxford University Press 2013.

\bibitem{agostinelli2003geant4}
S.~Agostinelli, et al.
\newblock GEANT4—a simulation toolkit.
\newblock {\em Nuclear instruments and methods in physics research section A: Accelerators, Spectrometers, Detectors and Associated Equipment}, 506(3):250--303, Elsevier 2003.

\bibitem{athanassas2020simulation}
C.~Athanassas, et al.
\newblock Simulation of a muographic analysis of a volcanic dome in Geant4.
\newblock {\em HNPS Advances in Nuclear Physics}, 27:37--47, 2020.

\bibitem{reyna2006simple}
D.~Reyna.
\newblock A simple parameterization of the cosmic-ray muon momentum spectra at the surface as a function of zenith angle.
\newblock {\em arXiv:hep-ph/0604145}, 2006.

\bibitem{kudryavtsev2009muon}
V.A.~kudryavtsev, et al.
\newblock Muon simulation codes MUSIC and MUSUN for underground physics.
\newblock {\em Computer Physics Communications}, 180(3):339--346, Elsevier 2009.

\bibitem{guardincerri2016imaging}
E.~Guardincerri, et al.
\newblock Imaging the inside of thick structures using cosmic rays.
\newblock {\em CAIP Advances}, 6(1):015213, AIP Publishing LLC 2016.

\bibitem{saracino2019applications}
G.~Saracino, et al.
\newblock Applications of muon absorption radiography to the fields of archaeology and civil engineering.
\newblock {\em Philosophical Transactions of the Royal Society A}, 377(2137):20180057, The Royal Society Publishing 2019.

\bibitem{menichelli2007scintillating}
M.~Menichelli, et al.
\newblock A scintillating fibres tracker detector for archaeological applications.
\newblock {\em Nuclear Instruments and Methods in Physics Research Section A: Accelerators, Spectrometers, Detectors and Associated Equipment}, 572(1):262--265, Elsevier 2007.

\bibitem{morishima2017discovery}
K.~Morishima, et al.
\newblock Discovery of a big void in Khufu’s Pyramid by observation of cosmic-ray muons.
\newblock {\em Nature}, 552(7685):386--390, Nature Publishing Group 2017.

\bibitem{Ambrosino_2015}
F.~Ambrosino, et al.
\newblock Assessing the feasibility of interrogating nuclear waste storage silos using cosmic-ray muons.
\newblock {\em Journal of Instrumentation}, 10(06):T06005--T06005, IOP Publishing 2015.

\bibitem{mahon2019first}
D.~Mahon, et al.
\newblock First-of-a-kind muography for nuclear waste characterization.
\newblock {\em Philosophical Transactions of the Royal Society A}, 377(2137):20180048, The Royal Society Publishing 2019.

\bibitem{schouten2019muon}
D.~Schouten.
\newblock Muon geotomography: selected case studies.
\newblock {\em Philosophical Transactions of the Royal Society A}, 377(2137):20180061, The Royal Society Publishing 2019.

\bibitem{bonneville2019borehole}
A.~Bonneville, et al.
\newblock Borehole muography of subsurface reservoirs.
\newblock {\em Philosophical Transactions of the Royal Society A}, 377(2137):20180060, The Royal Society Publishing 2019.

\bibitem{kudryavtsev2012monitoring}
V.A.~Kudryavtsev, et al.
\newblock Monitoring subsurface CO2 emplacement and security of storage using muon tomography.
\newblock {\em International journal of greenhouse gas control}, 11:21--24, Elsevier 2012.

\bibitem{gluyas2019passive}
J.~Gluyas, et al.
\newblock Passive, continuous monitoring of carbon dioxide geostorage using muon tomography.
\newblock {\em Philosophical Transactions of the Royal Society A}, 377(2137):20180059, The Royal Society Publishing 2019.

\bibitem{tanaka2009cosmic}
H.~Tanaka, et al.
\newblock Cosmic ray muon imaging of magma in a conduit: Degassing process of Satsuma Iwojima Volcano, Japan.
\newblock {\em Geophysical Research Letters}, 36(1), Wiley Online Library 2009.

\bibitem{LesparreMuon}
N.~Lesparre, et al.
\newblock Geophysical muon imaging: feasibility and limits.
\newblock {\em Geophysical Journal International}, 183(3):1348-1361, 2010.

\bibitem{tanaka2019japanese}
H.~Tanaka.
\newblock Japanese volcanoes visualized with muography.
\newblock {\em Philosophical Transactions of the Royal Society A}, 377(2137):20180142, The Royal Society Publishing 2019.

\bibitem{moss2018muon}
H.~Moss, et al.
\newblock Muon tomography for the cerro mach{\i}n volcano.
\newblock {\em Technical report, Department of Physics \& Astronomy, University of Sheffield}, 2018.

\bibitem{rodriguez2018minimute}
H.~Asorey, et al.
\newblock miniMuTe: A muon telescope prototype for studying volcanic structures with cosmic ray flux.
\newblock {\em Scientia et technica}, 23(3):386--391, 2018.

\bibitem{Pe_a_Rodr_guez_2020}
J.~Pe{\~{n}}a-Rodr{\'{\i}}guez, et al.
\newblock Design and construction of {MuTe}: a hybrid Muon Telescope to study Colombian volcanoes.
\newblock {\em Journal of Instrumentation}, 15(09):P09006--P09006, IOP Publishing 2020.

\bibitem{rosas2018muon}
M.~Rosas-Carbajal, et al.
\newblock Muon tomography of the La Soufri{\`e}re de Guadeloupe hydrothermal system: 3-D structure and dynamics.
\newblock {\em European Geosciences Union General Assembly 2018}, 20:EGU2018-18548, 2018.

\bibitem{varga2020tracking}
D.~Varga, et al.
\newblock Tracking detector for high performance cosmic muon imaging.
\newblock {\em Journal of Instrumentation}, 15(05):C05007, IOP Publishing 2020.

\bibitem{vasquez2020simulated}
A.~V{\'a}squez-Ram{\'\i}rez, et al.
\newblock Simulated response of MuTe, a hybrid Muon Telescope.
\newblock {\em Journal of Instrumentation}, 15(08):P08004, IOP Publishing 2020.

\bibitem{AsoreyNunezSuarez2018}
H.~Asorey, et al.
\newblock Preliminary results from the latin american giant observatory space weather simulation chain.
\newblock {\em Space Weather}, 16(5):461--475, Wiley Online Library 2018.

\bibitem{sarmiento2019modeling}
C.~Sarmiento-cano, et al.
\newblock Modeling the LAGO's detectors response to secondary particles at ground level from the Antarctic to Mexico.
\newblock {\em 36th International Cosmic Ray Conference}, 358(412), PoS 2019.

\bibitem{ambrosi2011mu}
G.~Ambrosi, et al.
\newblock The MU-RAY project: Volcano radiography with cosmic-ray muons.
\newblock {\em Nuclear Instruments and Methods in Physics Research Section A: Accelerators, Spectrometers, Detectors and Associated Equipment}, 628(1):120--123, Elsevier 2011.

\bibitem{nagamine1995method}
K.~Nagamine, et al.
\newblock Method of probing inner-structure of geophysical substance with the horizontal cosmic-ray muons and possible application to volcanic eruption prediction.
\newblock {\em Nuclear Instruments and Methods in Physics Research Section A: Accelerators, Spectrometers, Detectors and Associated Equipment}, 356(2):585--595, Elsevier 1995.

\bibitem{tanaka2008radiographic}
H.~Tanaka, et al.
\newblock Radiographic imaging below a volcanic crater floor with cosmic-ray muons.
\newblock {\em American Journal of Science}, 308(7):843--850, American Journal of Science 2008.

\bibitem{vesga2021simulated}
A.~Vesga-Ram{\'\i}rez, et al.
\newblock Simulated annealing for volcano muography.
\newblock {\em Journal of South American Earth Sciences}, 109:103248, Elsevier 2021.

\bibitem{agusto2017crater}
M.R.~Agusto, et al.
\newblock The crater lake of Copahue volcano (Argentina): geochemical and thermal changes between 1995 and 2015.
\newblock {\em Geological Society of London, Special Publications}, 437(01):107--130, 2017.

\bibitem{lamberti2021soil}
M.C.~Lamberti, et al.
\newblock Soil CO2 flux baseline in Planch{\'o}n--Peteroa Volcanic Complex, Southern Andes, Argentina-Chile.
\newblock {\em Journal of South American Earth Sciences}, 105:102930, Elsevier 2021.

\bibitem{barnoud2019bayesian}
A.~Barnoud, et al.
\newblock Bayesian joint muographic and gravimetric inversion applied to volcanoes.
\newblock {\em Geophysical Journal International}, 218(3):2179--2194, Oxford University Press 2019.

\bibitem{benton2020optimizing}
C.J.~Benton, et al.
\newblock Optimizing geophysical muon radiography using information theory.
\newblock {\em Geophysical Journal International}, 220(2):1078--1094, Oxford University Press 2020.

\bibitem{pla2001low}
A.~Pla-Dalmau, et al.
\newblock Low-cost extruded plastic scintillator.
\newblock {\em Nuclear Instruments and Methods in Physics Research Section A: Accelerators, Spectrometers, Detectors and Associated Equipment}, 466(3):482--491, Elsevier 2001.

\bibitem{bellamy1993absolute}
E.H.~Bellamy, et al.
\newblock Absolute calibration and monitoring of a spectrometric channel using a photomultiplier.
\newblock {\em Nuclear Instruments and Methods in Physics Research Section A: Accelerators, Spectrometers, Detectors and Associated Equipment}, 339(3):468--476, North-Holland 1994.

\bibitem{chirikov2001precise}
I.E.~Chirikov-Zorin, et al.
\newblock Precise analysis of the metal package photomultiplier spectra.
\newblock {\em Nuclear Instruments and Methods in Physics Research Section A: Accelerators, Spectrometers, Detectors and Associated Equipment}, 461(3):587--590, Elsevier 2001.

\bibitem{PHDAlmela}
A.~Almela.
\newblock Detectores de Muones del Observatorio Pierre Auger: Dise{\~n}o y T{\'e}cnicas de Operaci{\'o}n de se Electr{\'o}nica.
\newblock {\em PhD thesis at Universidad Tecnol{\'o}gica Nacional, Argentina}, 2019.

\bibitem{dossi2000methods}
R.~Dossi, et al.
\newblock Methods for precise photoelectron counting with photomultipliers.
\newblock {\em Nuclear Instruments and Methods in Physics Research Section A: Accelerators, Spectrometers, Detectors and Associated Equipment}, 451(3):623--637, Elsevier 2000.

\bibitem{gilbert2008consequences}
J.S.~Gilbert, S.J. Stephen.
\newblock The consequences of fluid motion in volcanic conduits.
\newblock {\em Geological Society, London, Special Publications}, 307(1):1--10, Geological Society of London 2008.

\bibitem{haberland2009structure}
C.~Haberland, et al.
\newblock Structure of the seismogenic zone of the southcentral Chilean margin revealed by local earthquake traveltime tomography.
\newblock {\em Journal of Geophysical Research: Solid Earth}, 114(B1), Wiley Online Library 2009.

\bibitem{brenguier20073}
F.~Brenguier, et al.
\newblock 3-D surface wave tomography of the Piton de la Fournaise volcano using seismic noise correlations.
\newblock {\em Geophysical research letters}, 34(2), Wiley Online Library 2007.

\bibitem{pessel2003multiscale}
M.~Pessel, D. Gibert.
\newblock Multiscale electrical impedance tomography.
\newblock {\em Journal of Geophysical Research: Solid Earth}, 108(B1), Wiley Online Library 2003.

\bibitem{deroussi2009localization}
S.~Deroussi, et al.
\newblock Localization of cavities in a thick lava flow by microgravimetry.
\newblock {\em Journal of Volcanology and Geothermal Research}, 184(1-2):193--198, Elsevier 2009.

\bibitem{carbone2003combined}
D.~Carbone, et al.
\newblock Combined discrete and continuous gravity observations at Mount Etna.
\newblock {\em Journal of Volcanology and Geothermal Research}, 123(1-2):123--135, Elsevier 2003.

\bibitem{tiede2005modeling}
C.~Tiede, et al.
\newblock Modeling the density at Merapi volcano area, Indonesia, via the inverse gravimetric problem.
\newblock {\em Geochemistry, Geophysics, Geosystems}, 6(9):, Wiley Online Library 2005.

\bibitem{bonomi2020applications}
G.~Bonomi, et al.
\newblock Applications of cosmic-ray muons.
\newblock {\em Progress in Particle and Nuclear Physics}, 112:103768, Elsevier 2020.

%------------
\end{thebibliography}

\end{document}